\documentclass[10pt,letterpaper]{article}
\usepackage[top=0.85in,left=2.75in,footskip=0.75in]{geometry}

\usepackage{amsmath,amssymb}

\usepackage{changepage}

\usepackage[utf8x]{inputenc}

\usepackage{textcomp,marvosym}

\usepackage{cite}

\usepackage{nameref,hyperref}

\usepackage[right]{lineno}

\usepackage{microtype}
\DisableLigatures[f]{encoding = *, family = * }

\usepackage[table]{xcolor}

\usepackage{array}

\newcolumntype{+}{!{\vrule width 2pt}}

\newlength\savedwidth

\raggedright
\usepackage[aboveskip=1pt,labelfont=bf,labelsep=period,justification=raggedright,singlelinecheck=off]{caption}

\makeatletter
\renewcommand{\@biblabel}[1]{\quad#1.}
\makeatother

\usepackage{lastpage,fancyhdr,graphicx}
\usepackage{epstopdf}
\fancyheadoffset[L]{2.25in}
\fancyfootoffset[L]{2.25in}
\usepackage{subcaption}
\usepackage[noend]{algpseudocode}
\usepackage{algorithm}
\usepackage{algorithmicx}
\usepackage{soul}
\algrenewcommand\algorithmicrequire{\textbf{Input:}}
\algrenewcommand\algorithmicensure{\textbf{Output:}}
\renewcommand{\P}{{\mathbb P}}

\begin{document}
\vspace*{0.2in}

\begin{flushleft}
{\Large
\textbf\newline{Markovian city-scale modelling and mitigation of micro-particles from tires} % Please use "sentence case" for title and headings (capitalize only the first word in a title (or heading), the first word in a subtitle (or subheading), and any proper nouns).
}
\newline
\\
\def\Authors{, , , }
Gunda Obereigner\textsuperscript{1},
Roman Overko\textsuperscript{2*},
Serife Yilmaz\textsuperscript{3},
Emanuele Crisostomi\textsuperscript{4},
Robert Shorten\textsuperscript{5}
\\
\bigskip
\textbf{1} Institute for Design and Control of Mechatronical Systems, Johannes Kepler University, 
Linz, Austria
\\
\textbf{2} School of Electrical and Electronic Engineering, University College Dublin, Dublin, County 
Dublin, Ireland
\\
\textbf{3} Education Faculty, Department of Mathematics Education, Mehmet Akif Ersoy University, Burdur, 
Turkey
\\
\textbf{4} Department of Energy, Systems, Territory and Constructions Engineering, University of Pisa, 
Pisa, Italy
\\
\textbf{5} Dyson School of Design Engineering, Imperial College London, South Kensington, London, U.K.
\\
%Name1 Surname\textsuperscript{1,2\Yinyang},
%Name2 Surname\textsuperscript{2\Yinyang},
%Name3 Surname\textsuperscript{2,3\textcurrency},
%Name4 Surname\textsuperscript{2},
%Name5 Surname\textsuperscript{2\ddag},
%Name6 Surname\textsuperscript{2\ddag},
%Name7 Surname\textsuperscript{1,2,3*},
%with the Lorem Ipsum Consortium\textsuperscript{\textpilcrow}
%\\
%\bigskip
%\textbf{1} Affiliation Dept/Program/Center, Institution Name, City, State, Country
%\\
%\textbf{2} Affiliation Dept/Program/Center, Institution Name, City, State, Country
%\\
%\textbf{3} Affiliation Dept/Program/Center, Institution Name, City, State, Country
%\\
\bigskip

% Insert additional author notes using the symbols described below. Insert symbol callouts after author names as necessary.
% 
% Remove or comment out the author notes below if they aren't used.
%
% Primary Equal Contribution Note
%\Yinyang These authors contributed equally to this work.

% Additional Equal Contribution Note
% Also use this double-dagger symbol for special authorship notes, such as senior authorship.
%\ddag These authors also contributed equally to this work.

%\textcurrency Current Address: Dept/Program/Center, Institution Name, City, State, Country % change symbol to "\textcurrency a" if more than one current address note
% \textcurrency b Insert second current address 
% \textcurrency c Insert third current address


% Group/Consortium Author Note
%\textpilcrow Membership list can be found in the Acknowledgments section.

% Use the asterisk to denote corresponding authorship and provide email address in note below.
* roman.overko@ucdconnect.ie

\end{flushleft}
\section*{Abstract}

The recent uptake in popularity in vehicles with zero tailpipe emissions is a welcome development in the fight against traffic induced airborne pollutants. As vehicle fleets become electrified, and tailpipe emissions become less prevalent, non-tailpipe emissions (from tires and brake disks) will become the dominant source of traffic related emissions, and will in all likelihood become a major concern for human health. This trend is likely to be  exacerbated by the heavier weight of electric vehicles, their increased power, and their increased torque capabilities, when compared with traditional vehicles. While the problem of emissions from tire wear is well-known, issues around the process of tire abrasion, its impact on the environment, and modelling and mitigation measures, remain relatively unexplored. Work on this topic has proceeded in several discrete directions including:  on-vehicle collection methods; vehicle tire-wear abatement algorithms and controlling the ride characteristics of a vehicle, all with a view to abating tire emissions.  Additional approaches include access control mechanisms to manage aggregate tire emissions in a geofenced area with other notable work focussing on understanding the particle size distribution of tire generated PM, the degree to which particles become airborne, and the health impacts of tire emissions.  While such efforts are already underway, the problem of developing models to predict the 
aggregate picture of a network of vehicles at the scale of a city, has yet to be considered. Our objective in this paper is to present one such model, built using ideas from Markov chains. 
Applications of our modelling approach are given toward the end of this note, both to illustrate the utility of the proposed method, and to illustrate its application as part of a method to collect tire dust particles.

% Please keep the Author Summary between 150 and 200 words
% Use first person. PLOS ONE authors please skip this step. 
% Author Summary not valid for PLOS ONE submissions.   
%\section*{Author summary}

%\linenumbers

% Use "Eq" instead of "Equation" for equation citations.
% For figure citations, please use "Fig" instead of "Figure".
% Place tables after the first paragraph in which they are cited.
% Place figure captions after the first paragraph in which they are cited.
% Results and Discussion can be combined.
% PLOS does not support heading levels beyond the 3rd (no 4th level headings).

\section{Introduction}
The recent uptake in popularity in vehicles with zero tailpipe emissions is a welcome development in 
the fight against traffic induced airborne pollutants. The deployment of such vehicles is consistent 
with the prevailing contemporary narrative which is heavily focussed on mechanisms to abate mobility 
related  greenhouse gases and tailpipe pollutants; see~\cite{shorten1} for a snapshot of some recent 
work across several disciplines on this topic. However, as vehicle fleets become electrified, 
non-tailpipe emissions (from tires and brake disks) are likely to become a major concern for human 
health and this is likely to be exacerbated by the transition to electric vehicles due to their heavier 
weight and increased torque capabilities~\cite{shorten3,austria_again}.
\newline

The issue of emissions from tire wear is in itself a very old topic. Somewhat remarkably, issues 
around the process of tire abrasion, its impact on the environment and human health, and modelling 
and mitigation measures, remain relatively unexplored and poorly understood. In addition, the general
public seems oblivious to the fact that these emissions are significant and almost certainly harmful to 
human health. The fact that the topic is 
relatively unexplored and unknown (by the general public) in the context of automotive engineering is very surprising given the rate 
at which tire mass abrades and contributes to particulate matter (PM) in moving vehicles. PM is a 
generic term used for a type of pollutants that consists of a complex and varied mix of small particles. 
There is a growing and rich literature documenting the link between PM and its effects on human 
health~\cite{valavanidis,gehring,eea2014,harvard_2006,air_survey_2014,pm_brazil_2011}. A recent review of the impact of tire and road wear particles can be found in~\cite{Baltruschat_2020}. 
Roughly speaking, smaller PM particles tend to be directly more harmful to humans compared to larger ones, 
as they can travel deeper into the respiratory system~\cite{dementia,toronto_2010,who,eea2014} (though larger 
toxic particles can also cause harm if they enter our food systems). 
Some of the known health effects related to PM include oxidative stress, inflammation and early atherosclerosis. 
Other studies have shown that smaller 
particles may go into the bloodstream and thus translocate to the liver, the kidneys or the brain 
(see~\cite{non_exhaust} and references within).  
\begin{figure}[!h]
    \begin{center}
        \includegraphics[width=\textwidth]{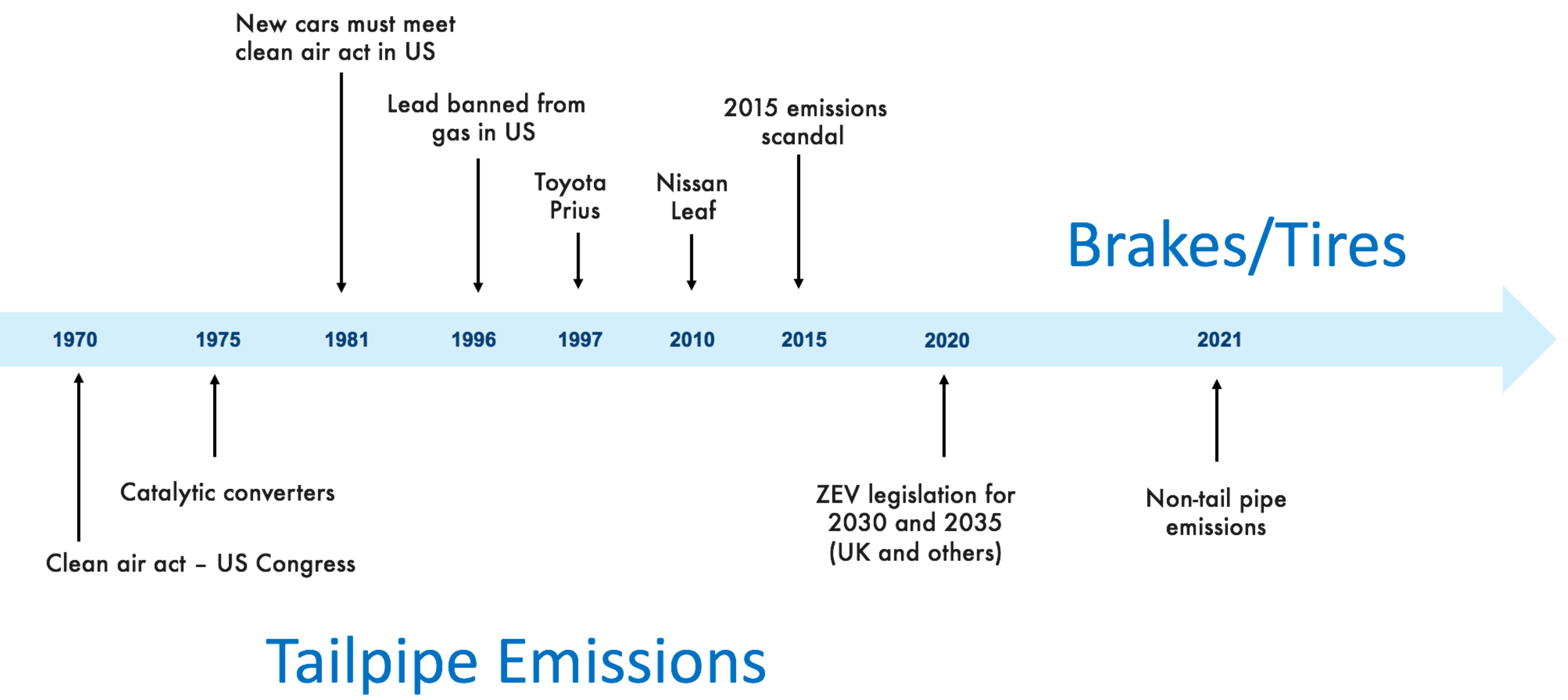}
    \end{center}
    \caption{Evolution of recent vehicle emissions narrative, and emerging non-tailpipe emissions concerns.}
    \label{fig:history}
\end{figure}
According to the World Health Organisation, for 
$P\!M_{2.5}$, the daily maximum deemed safe level on average is $25\ \mu g/m^3$, whereas the annual 
maximum permitted level is on average $10\ \mu g/m^3$. For $P\!M_{10}$, the maximum permitted levels 
are on average $50\ \mu g/m^3$ and $20\ \mu g/m^3$ on a daily and annual basis, respectively. 
In general, based on these numbers, it is acknowledged that non-exhaust emissions (including brake and 
tire wear, road surface wear and resuspension of road dust) resulting from road traffic, account for 
a significant component of traffic related PM emissions~\cite{non_exhaust_electric}. 
To parse these numbers in the context of a specific city it was recently estimated that approximately 
186 kg per day of tire mass is lost to abrasion in Dublin each day~\cite{shorten3}.\newline 

Recently, the issue 
of tire generated PM emissions has become a topic of interest for several groups 
worldwide~\cite{microplastics, grigoratos,netherlands_2010}. Roughly speaking, work on this topic 
has proceeded in several directions with work on the topic focussing on on-vehicle collection methods, 
on vehicle tire-wear abatement algorithms, or estimating properties of tire debris. For example, 
several of the authors of~\cite{shorten3}, {\em the tire Collective}, have constructed a prototype 
on-wheel system for collection of tire 
debris\protect\footnote{https://www.youtube.com/watch?v=fo-2b5JzTl8}. Other authors~\cite{gunda1,gunda2} 
have explored controlling the ride characteristics of a vehicle with a view to abating tire emissions. 
A further approach in~\cite{shorten3} explores access control mechanisms to manage aggregate tire 
emissions in a geofenced area. Other notable work on the topic focusses on the particle size 
distribution of tire generated PM, or to which degree this becomes airborne~\cite{air_survey_2014} 
(while currently available emission factors for tire wear in literature gives estimates of vehicle emissions 
of between $0.005-100g/km$, no reliable method to calculate tire related PM or tire wear, depending, 
for example, on the vehicle operation, appears to be available~\cite{fausser,non_exhaust_electric}). 
The issue of which particulates become airborne is in fact the subject of some debate in the community.
We note strongly that we are {\bf not} concerned with such classifications. While research on emissions has focussed on airborne pollutants, 
the reality is that both outcomes are problematic for humans. Particles that become 
airborne have the potential to contribute to poor air quality in cities with all the ensuing health 
consequences; those that fall to the ground have the potential to enter water systems and contribute 
to the general problem of environmental microplastic pollution. Thus both manifestations of the 
tire pollution problem need to be addressed.\newline  

Our objective in this paper is to 
develop city-scale models of tire pollution, for both airborne and non-airborne PM, that can be used to 
inform policy makers in the fight to mitigate the effect of tire abrasion. We have already mentioned 
that the issue of tire wear is an old and relatively unexplored topic,in automotive engineering,
and is subject to sources of large uncertainty. For example, tire induced PM, depends not only on 
the chemical composition of tires, but also on traffic densities, speeds, driving styles, and road surfaces. 
Indeed, the ultimate impact on humans depends on the effect of large aggregations of vehicles,
each driven by drivers with differing styles, and with different tires. Given this uncertainty, 
there is clearly a need for robust and efficient methods that indicate the likely locations where large accumulations of tire 
mass are likely to be found. We would also like that these models somehow capture the complex relationship between 
speed limits, traffic signalling, and densities, so that these parameters can be explored from the perspective of tire emissions. 
To do this we shall build on our previous work on Markovian~\cite{Crisostomi2011} models of traffic networks. 
An important point to note in this context is that even though tire emission factors are not well known (perhaps even 
unknowable), the qualitative aspects of the tire abrasion process is understood (the qualitative 
effects of speed, acceleration, weight, road surface). This is important from the context of Markovian 
network emission models which, even though uncertain, do tell us where build ups are likely to occur 
and the importance of road segments from the context of road debris.    We shall show how such models 
enable a number of important applications in the fight against tire dust; in particular how such models 
can be used to inform tire dust collection strategies and to inform vulnerable road users such as 
pedestrians and cyclists.\newline

\section{Markovian models of traffic systems}
The use of Markov chains for traffic congestion analysis was first proposed in~\cite{Crisostomi2011}. 
Since then the idea has been developed and applied to other traffic related issues in a series of 
papers~\cite{Chapter2011, Multimodal} and by other authors~\cite{Others1, Others2}. For convenience, 
we now briefly recall some of the background discussion on such models, while a more thorough 
explanation on such models can be found in the previous references. Traffic flows can be described 
through a \textit{Markov chain}, which is a stochastic process characterized by the equation 
\begin{equation}
\label{eq:1}
p(x_{k+1} = {S_i}_{k+1}|x_k = {S_i}_k, \dots, x_0 = {S_i}_0) = p(x_{k+1} = {S_i}_{k+1}|x_k = {S_i}_k ) 
\quad \forall k\geq 0, 
\end{equation}
where $p(E|F)$ denotes the conditional probability that event $E$ occurs given that event $F$ occurs. 
Eq~(\ref{eq:1}) states that the probability that the random variable $x$ is in state ${S_i}_{k+1}$ 
at time step $k+1$ only depends on the state of $x$ at time step $k$ and not on preceding values. 
Usually the Markov chain with $n$ states is described by the \textit{$n \times n$ transition probability 
matrix} $\P$, whose entry $\P_{ij}$ denotes the probability of passing from state $S_i$ to state $S_j$ 
in exactly one step. Clearly the matrix $\P$ is a matrix whose rows sum to one (row-stochastic 
non-negative matrix).
\newline

Markov chains are particularly useful for traffic systems due to their close association with graphs 
(in the context of traffic road networks). Recall that a graph is represented by a set of nodes that 
are connected through \textit{edges}. Therefore, the graph associated with the matrix $\P$ is a 
\textit{directed graph}, whose nodes are represented by the states $S_i$, $i=\{1,\ldots, n\}$ and 
there is a directed edge leading from $S_i$ to $S_j$ if and only if $P_{ij}\neq0$. The strong connection 
between graphs and Markov chains manifests itself in many ways. For example, the notions of chain 
irreducibility and strongly connected graphs are enunciations of the same concept. More precisely, a 
graph is \textit{strongly connected} if and only if for each pair of nodes there is a sequence of 
directed edges leading from the first node to the second one. Thus, $\mathbb{P}$ is \textit{irreducible} 
if and only if its directed graph is strongly connected. The usefulness of Markov chains for road 
networks extends well beyond their close relation to graphs. In particular, many easily computable 
properties of the chain (from the transition matrix) also have strong physical interpretations. For 
example, for irreducible transition matrices, it is known that the {\em spectral radius} of $\P$ is $1$. 
This fact is used in applications to detect communities in chains associated with transportation 
networks. Moreover, the left-hand Perron eigenvector $\pi$ of the $\P$ matrix, that is $\pi^T P = \pi^T$ 
such that $\pi_i>0$, $||\pi||_1=1$, yields a closed form expression for the stationary distribution of 
a random walker over the graph associated with the Markov chain. As such it has a strong connection to 
likely congestion locations in transportation networks. We shall exploit the Perron eigenvector in the 
present paper for the purpose of determining likely locations of high tire emissions. Finally, two 
other quantities that are useful for studying graphs and which can be easily computed are the 
{\em Kemeny constant} and the {\em Mean First Passage Time}. The mean first passage time (MFPT) 
$m_{ij}$ from the state $S_i$ to the state $S_j$ denotes the expected number of steps to arrive at 
destination $S_j$ when the origin is $S_i$, and the expectation is averaged over all possible paths 
following a random walk from $S_i$ to $S_j$.  If we assume that $m_{ii}=0$, then the 
\textit{Kemeny constant} is defined as
\begin{eqnarray}
\label{Kemeny_Equation}
K= \sum_{j=1}^n m_{ij} \pi_j.
\end{eqnarray}
Remarkably, the right-hand side is independent of the choice of the origin state $S_i$~\cite{Kemeny1960}. 
An interpretation of this result is that the expected time to get from an initial state $S_i$ to a 
destination state $S_j$ (selected randomly according to the stationary distribution $\pi$) does not 
depend on the starting point $S_i$~\cite{doyle2009kemeny}. Therefore, the Kemeny constant is an 
intrinsic measure of a Markov chain. Eq~(\ref{Kemeny_Equation}) emphasizes the fact that $K$ is only 
related to the particular matrix $\P$ and it becomes very large if one or more of the other eigenvalues 
of $\P$, different from $\lambda_1$, are close to $1$.\newline

The use of Markov chains to model road network dynamics has been described in detail 
in~\cite{Crisostomi2011} and in many subsequent papers by other authors~\cite{Others1, Others2}. 
The resulting networks are fully characterized by a transition matrix $\P$, which has the following form:
\begin{equation}
\label{trans_matrix_traffic}
   \P=
  \left[ {\begin{array}{cccc}
   P_{S_1 \to S_1} & P_{S_1 \to S_2} & \cdots & P_{S_1 \to S_n} \\
   P_{S_2 \to S_1} & P_{S_2 \to S_2} & \cdots & P_{S_2 \to S_n} \\
   \vdots          & \vdots          & \ddots & \vdots \\
   P_{S_n \to S_1} & P_{S_n \to S_2} & \cdots & P_{S_n \to S_n} \\
 \end{array} } \right].
\end{equation}
The matrix $\P$ is a square matrix whose size is given by the number of road segments. The off-diagonal 
elements ${\P}_{S_i \to S_j}$ are related to the probability that one passes directly from the road 
segment $S_i$ to the road segment $S_j$. Importantly, the transition matrix can be very easily computed 
after gathering the average travel times and junction turning probabilities. In our models the diagonal 
terms are proportional to travel times. If travel times are computed for all roads, and they are 
normalized so that the smallest travel time is 1, then the probability value associated to each 
self-loop is
\begin{eqnarray} 
\label{diagonaltravel}
P_{S_i \to S_i}= \frac{tt_i-1}{tt_i}, \,\ i=\{1,\ldots,n \}
\end{eqnarray}
where $tt_i$ is the average travel time (estimated from collected data) for the $i$-th road. The 
off-diagonal elements of the transition matrix $\P$ can be obtained as
\begin{eqnarray} 
\label{offdiagonaltravel}
P_{S_i \to S_j}= (1- P_{S_i \to S_i})\cdot(t p_{ij}) \,\ i \neq j,
\end{eqnarray}
where $tp_{ij}$ is the turning probability (estimated from collected data) of going from road $i$ to 
road $j$~\cite{Crisostomi2011}. In the next section we shall explain how this basic transition 
matrix~(\ref{trans_matrix_traffic}) can be modified to convert the model the evolution of tire 
emissions in an urban landscape.\newline

{\bf Comment:} The interested reader may ask the advantage of a Markovian model
of traffic, as compared with using a traffic simulator, such as SUMO, which we shall extensively use in the remainder of this paper for validation purposes. The principal advantages of a theoretical approach are 
threefold. First, in terms of utility, once identified, the Markovian model gives access to predictions 
in a very efficient manner, especially when compared with Monte-Carlo based approached based 
on vehicle simulators. Second, following from the previous point, the parameters of mathematical 
models can be efficiently adjusted to explore traffic management strategies, without the need for 
ensembles of complex simulations. Finally, by developing a Markovian (transition matrix) approach
to traffic modelling, one may avail of a well developed suite of analytics that have been developed to analyse
Markovian systems over the past century. This can then be used both to study and analyse the properties 
of transportation networks, as well as providing a basis for the design of network-level traffic policies. Indeed,
this has been explored in a series of papers on traffic modelling since the publication of \cite{Crisostomi2011}; see \cite{Multimodal} for
examples of work in this direction. \newline

\section{Extension of Markovian Model to tire Emissions}\label{tire_Emissions_Model}%TODO
Our starting point in developing a tire emissions model is the assumption 
that the Perron eigenvector of a traffic congestion matrix also provides 
some relevant information about tires' emissions. This is a reasonable assumption because the entries of the Perron eigenvector 
report the average long-run fraction of time that a vehicle spends on each road. However, there is not 
a precise relationship between emissions and density information as tires emissions do not only depend on the 
amount of time that is spent along one road, but also on other quantities, such as average driving style and 
average speeds. To capture such effects, as a first 
approximation to develop city-scale models of tire pollution, we shall now describe how the number of 
tire particles can be estimated depending on the vehicle's speed, and how this information can be 
embedded in the Markov chain transition matrix.\newline

As we have mentioned, tire emission factors in the literature are characterised by huge uncertainty 
varying between $0.005$ to $100 g/km$~\cite{non_exhaust}. In any case such a simple characterisation of 
tire based PM is not suitable to build a Markovian model of tire emissions; to build such a model a 
tire based PM estimation model depending on a vehicle's operation mode (for example, speed, acceleration, 
driving style) is required. To this end, we shall use measurements and results from~\cite{Foitzik} that show a dependency of the 
number of ultra-fine tire particles $PN$ produced by a vehicle and the vehicle's operation. As particles irrespective of size, 
tend to be harmful to human health~\cite{Schraufnagel2} we shall in the sequel focus on the number of particles to 
evaluate the impact in the city network, and consequently adopt and develop the approach from~\cite{Foitzik} 
to estimate the number of particles.\newline 

{\bf Comment:} To further justify our approach it is worth noting the approach adopted here has 
also been recognised by the latest EU legislative regulations. These place a higher emphasis on the number of particles rather than particle mass or size 
distribution~\cite{EU_regulation}.\newline 

The measurements from~\cite{Foitzik} show a linear 
dependency of $PN$ and vehicle's speed $v$, as well as an approximately quadratic dependency of $PN$ 
and the forces on the wheels $F$. The combination of both curves leads to the following estimate of $PN$
\begin{equation}
\label{eq: PN(F,v)}
PN(F,v) = (a_0 +a_1 \cdot F + a_2 \cdot F^2)\cdot (b_0 + b_1 \cdot v).
\end{equation}
It is not a trivial matter to gather information about values of $F$ and it is therefore more convenient to express 
$F$ as a function of $v$ (as aggregate estimates of $v$ are simpler to obtain). To do so, we make the 
simplifying assumption that all roads in the city network are flat (without incline or elevation) and,  in addition, for the sake of simplicity, 
accelerations are neglected.\newline 

{\bf Comment:} The previous assumption introduces some approximations in our estimates 
(for example, accelerations would cause a higher number of tire particles~(\ref{eq: PN(F,v)})). However, we make two observations. 
\begin{itemize}
\item[(i)] First, it is important to 
note that if city-wide accelerations can be measured and aggregated, then this approach can be corrected to yield a more realistic and 
sophisticated model for tire particle estimation.
\item[(ii)] In many of our applications, we are interested in locations of elevated tire dust. While the simplified modelling approach will certainly affect the 
estimate of absolute amount of PM gathered in a specific location, the relative ranking of locations (to guide, for example, collection) will be less affected 
by the modelling assumption.\newline 
\end{itemize}
Thus, in our case, the force can be 
approximated as a function of velocity as
\begin{equation}
\label{eq: F(v)}
F(v) = m \cdot g \cdot c_\text{r} + \frac{\rho A c_\text{d}}{2}v^2,
\end{equation}
where the first term describes the rolling effect of the vehicle, while the second term takes into account the air 
drag resistance. In Eq~(\ref{eq: F(v)}), $m$ is the mass of the vehicle, $g$ is the the gravitational 
constant, $c_\text{r}$ is the rolling resistance, $c_\text{d}$ is the drag resistance coefficient, $\rho$ 
is the density of air, and $A$ is the approximated front area of the vehicle. Numerical values for an average vehicle 
are given in Table~\ref{tab: PN(v) parameter}. To estimate the number of tire particles per driven 
km,~(\ref{eq: PN(F,v)}) and~(\ref{eq: F(v)}) are combined as follows: 
\begin{equation}
\label{eq: PN(v)}
PN(v) = (a_0 +a_1 \cdot \frac{F(v)}{1000} + a_2 \cdot \left(\frac{F(v)}{1000}\right)^2)\cdot (b_0 + b_1 
\cdot v) \cdot \frac{1000}{v}.
\end{equation}
Note that the true process of generating tire particles is also affected by other factors that we are 
not considering here, such as road surface, type of tire, vehicle's weight~\cite{Jekel} etc.\newline
 
{\bf Comment :} As a final 
comment, we further remark that Eq~(\ref{eq: PN(v)}) gives the number of particles under rather 
approximated conditions and may underestimate the actual number of $PN$, as acceleration and braking 
events are neglected. However, as the Markovian models reveal densities, we expect these approximations 
to be reasonably accurate up to a scaling factor.\newline 

\begin{table}[!ht]
    \centering
    \caption{Numerical values for parameters in $PN(v)$~(\ref{eq: PN(v)}).}\vspace{0.25cm}
    \begin{tabular}{|ccc|}
        \hline
        Parameter & Value & Unit      \\
        \hline
        $m$         & 2200    & kg    \\
        $g$         & 9.81     & m/s$^2$   \\
        $c_\text{r}$        & 0.0108 &  -   \\
        $\rho$       & 1.2      & kg/m$^3$  \\
        $A $        & 2.2      & m$^2$   \\
        $c_\text{d}$        & 0.233  & -    \\
        \hline
    \end{tabular}
    \label{tab: PN(v) parameter}
\end{table}

The estimate of the number of tire particles~(\ref{eq: PN(v)}) is now used to convert the unit of 
time of the original transition matrix into a tire particle, and a step in the Markov chain now 
corresponds to a unit of tire emission. For this purpose, we change the diagonal entries of the 
transition matrix $\P$ as follows:
\begin{equation}
\label{Diagonal_Terms}
P_{S_i \to S_i}= \frac{PN(v_i) \cdot l_i-1}{PN(v_i)\cdot l_i}, \,\ i=\{1,\ldots,n \},
\end{equation}
where $v_i$ is the average velocity on the road segment $i$ and $l_i$ is the length of the corresponding 
road segment. Then, the off-diagonal elements are re-normalized as stated in Eq~(\ref{offdiagonaltravel}) 
to keep the transition matrix row-stochastic.\newline
 
{\bf Comment :} The effect of the diagonal scaling of Eq~(\ref{eq: PN(v)}) is that large values in the diagonal of the original matrix 
$\P$ of Eq~(\ref{eq: PN(v)}) corresponded to long times required to travel along a given road segment, while in the new transition matrix for the tire emissions model, large values in the diagonal entries of the new transition matrix now correspond to road segments with high tire emissions. More properties of the diagonal scaling can be found in~\cite{Chapter2011}. Table \ref{Table_Comparison} summarizes the interpretation of typical quantities of interest in Markov chains for the original transition matrix of travel times, and the new transition matrix related to tire emissions.\newline

\begin{table}[!ht]
\centering
\caption{Interpretation of some Markov chain quantities of interest in: (a) the case of the Markov chain characterising road congestion; (b) the case of the Markov chain characterising tire emissions.\\}\vspace{0.25cm}
{\begin{tabular}{|p{.25\textwidth}||p{.3\textwidth}|p{.3\textwidth}|}
\hline
\textbf{Quantity $/$ Markov chain} & \textbf{Congestion} & \textbf{Tire emissions}\\
\hline
\textbf{Perron Eigenvector} & Vehicular density in the network & Density of tires emissions in the network\\
\hline
\textbf{Mean First Passage Times} & Average travel times for a pair of origin/destination & Average amount of emissions for a pair of origin/destination\\
\hline
\textbf{Kemeny constant} & Average travel time for a random trip (Global indicator of travel efficiency) & Average amount of emissions for a random trip (Global indicator of tires-induced emissions) \\
\hline
\end{tabular}}
\label{Table_Comparison}
\end{table}

\section{Applications of tire Emissions Model}\label{Interpretation}%TODO
While the utility of Markovian traffic models has been documented in several publication, to the best of our knowledge their 
utility in the context of tire emissions has not yet been investigated. The objective of this section is to present some basic applications of the 
tire emissions model to illustrate its utility. We begin with some basic applications.\newline

\subsection{Application I - Design of Low Emissions Zones}
\begin{figure}[!h]
    \begin{center}
        \includegraphics[width=\textwidth]{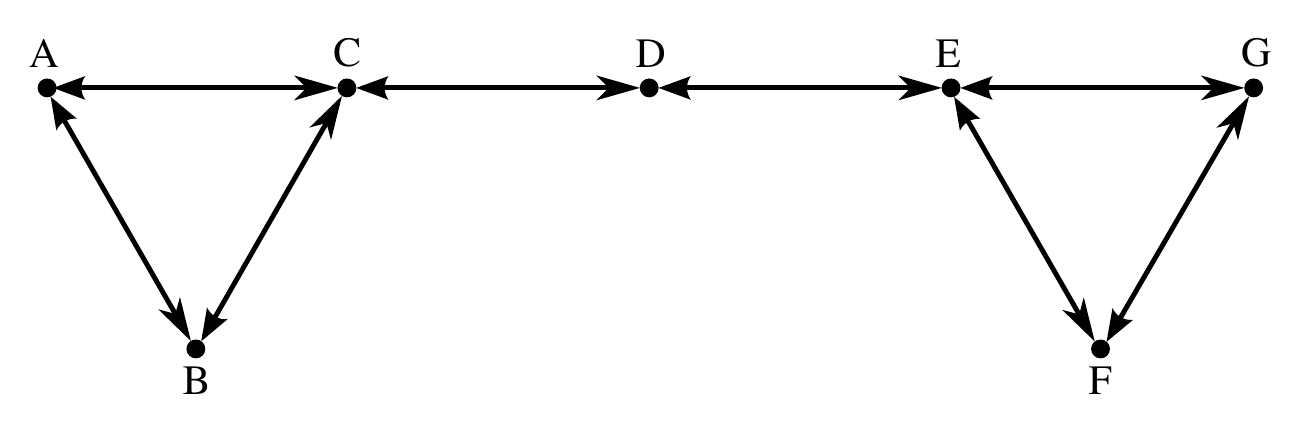}
    \end{center}
    \caption{Simple urban network.}
    \label{fig: network_simple}
\end{figure}

To illustrate potential 
applications of our approach, we now consider the design of a low emissions zone for a city. To provide some background context and link this to our previous work we now first consider the same urban network that had 
been investigated in references~\cite{Crisostomi2011,Chapter2011}. This simple network is depicted in 
Fig~\ref{fig: network_simple} and assumes that two clusters of nodes A-B-C and E-F-G are connected 
through node D. In the diagram nodes correspond to junctions, and links to roads. 

\begin{figure}[!h]
    \begin{center}
        \includegraphics[width=\textwidth]{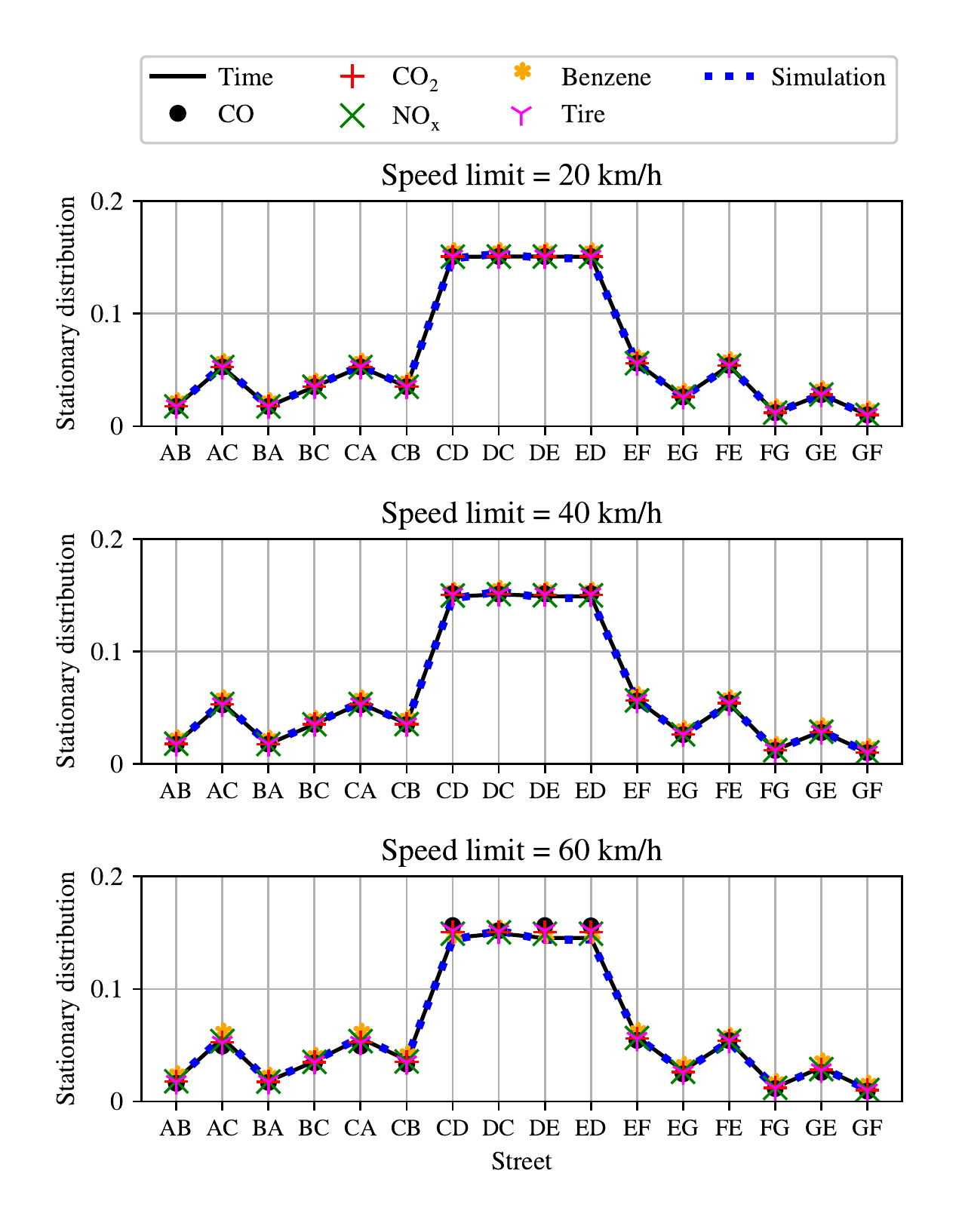}
    \end{center}
    \caption{Distribution of emissions for the simple network with different speed limits,
        namely $20$, $40$ and $60 km/h$, in the entire network.}
    \label{fig:network_simple_emissDist_speed_everywhere_a}
\end{figure}

\begin{figure}[!h]
    \begin{center}
        \includegraphics[width=\textwidth]{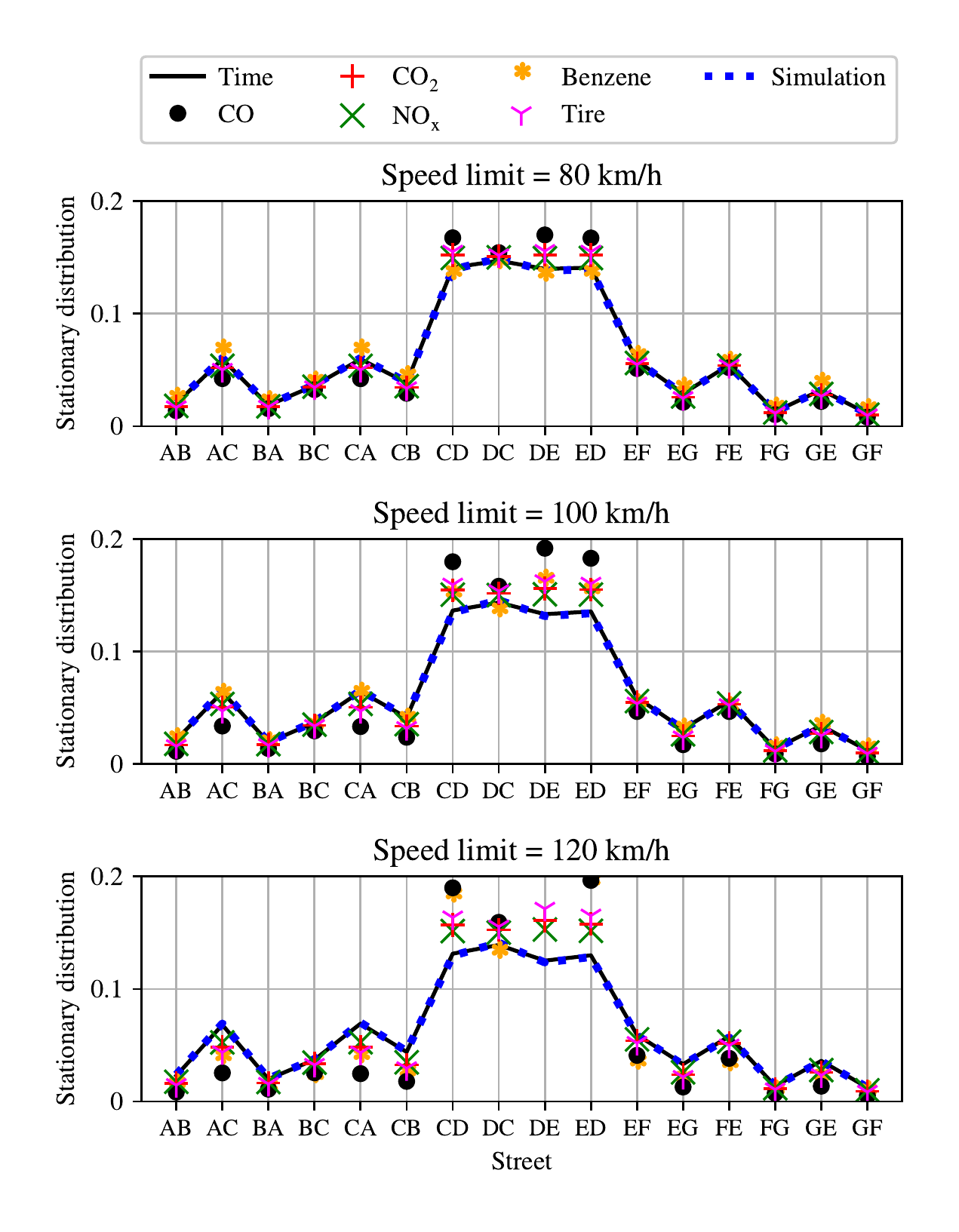}
    \end{center}
    \caption{Distribution of emissions for the simple network with different speed limits,
    namely $80$, $100$ and $120 km/h$, in the entire network.}
    \label{fig:network_simple_emissDist_speed_everywhere_b}
\end{figure}

It is well known that changing speed limits may be a convenient policy to reduce urban emissions of 
pollutants. To investigate this idea for tire emissions we now compare what happens if different speed limits are considered in the whole network, 
and results are illustrated in Fig~\ref{fig:network_simple_emissDist_speed_everywhere_a}, 
\ref{fig:network_simple_emissDist_speed_everywhere_b}, in terms of the 
entries of the Perron eigenvector of the stationary distribution of tire emissions. Recall, the entries 
of the Perron vector are those road segments where tire emissions are most likely to accumulate (both 
airborne and on the ground).\newline

As it can be seen from Fig~\ref{fig:network_simple_emissDist_speed_everywhere_a}, 
\ref{fig:network_simple_emissDist_speed_everywhere_b}, measured vehicle obtained 
from the mobility simulator SUMO\footnote{\url{https://www.eclipse.org/sumo/}} (blue dashed line) are 
compared with the basic Markovian traffic model (black solid line), for every considered speed limit. 
A very close correspondence between the simulator output and the Markov chain can be observed. In the 
same figure, we also report the distribution of different pollutants as estimated using the Markov 
chain of emissions~\cite{Chapter2011}.\newline
 
{\bf Comment :} While the stationary distributions of different pollutants and travel times can be easily estimated with the Markovian approach in a few milliseconds, it is more complicated to retrieve the same information by using the simulator. Indeed, in the latter case an ensemble of simulations has to be carried out for each different value of speed limits, to average the stochastic effects of different routes of different vehicles.\newline

\subsubsection{Optimized speed limits}
\begin{figure}[!h]
    \begin{center}
        \includegraphics[width=\textwidth]{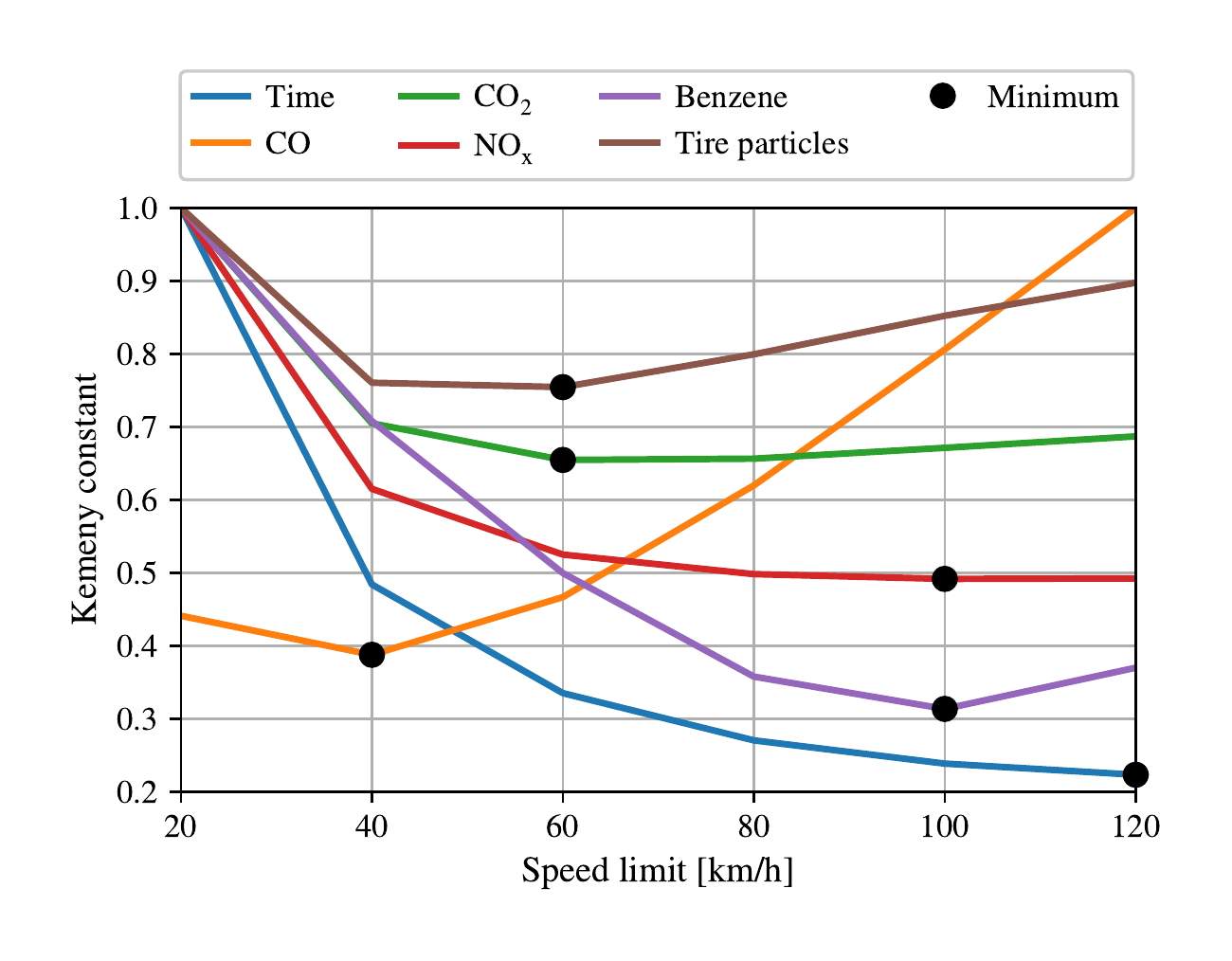}
    \end{center}
    \caption{Kemeny constants for the simple network with different speed limits in the entire network.}
    \label{fig:network_simple_kemeny_speed_everywhere}
\end{figure}

It can be observed in Fig~\ref{fig:network_simple_emissDist_speed_everywhere_a} that for low speed limits the density of pollutants is proportional to the density of vehicles (where there are more vehicles, there is more pollution), however, when higher speed limits are considered, the proportionality is lost, and different pollutants exhibit different properties with different speed limits.
This last comment is further illustrated in Fig~\ref{fig:network_simple_kemeny_speed_everywhere} that shows the optimal 
value of the Kemeny constant as a function of speed limits. 
Recall that in the context of the tire emissions model, the Kemeny constant $K$ is a measure of 
the average number of emissions associated with trips in the networks, and thus. it is a single quantity 
of the Markov chain which can be interpreted as an indication for the pollution in the entire network.\newline

In particular, a lower value the Kemeny constant corresponds to a lower value of average emissions and a better overall network. As we have 
already observed, this may however be a tricky problem, 
since the optimal  speed limits for tire particles may actually increase emissions from other pollutants. In order to 
calculate the optimal speed limit for the network, simulations for different maximum speeds, which are 
$20, 35, 50,65, 80,95$\,km/h, have been conducted using SUMO. After simulating the network for these six 
different speed limits, and building the resulting Markov chain, six different Kemeny constants can be 
calculated for each type of pollutant as well as for the travel time. It can be clearly observed that the optimal 
speed limit varies again for different types of pollutants. While low speed limits seem to be good for 
reducing $CO$ and tire particles, high speed limits would be better to reduce $NOx$ and of course travel 
time. Figure ~\ref{fig:network_simple_kemeny_speed_everywhere} depicts the non-obvious 
result that the ``environmentally optimal'' speed limit actually depends on the specific pollutant that 
one is interested in minimizing. In particular, $40 km/h$ appears to be the best speed limit if one aims at 
minimizing $CO$ emissions, $60 km/h$ is the best choice for minimizing $CO_2$ and tire emissions (which 
is the specific objective of this manuscript), $100 km/h$ is the best choice for minimizing $NOx$ and 
Benzene, while, obviously, the maximum considered speed limit (i.e., $120 km/h$) is the best option to 
minimize travel times. Thus, the selections of the ``best'' speed limits is not trivial, and policy makers should be 
informed about the optimal speed limits for different pollutants in order to make a decision.\newline

\subsubsection{Optimal speed limits in more realistic road networks}

\begin{figure}[!h]
    \begin{center}
        \includegraphics[width=0.7\textwidth]{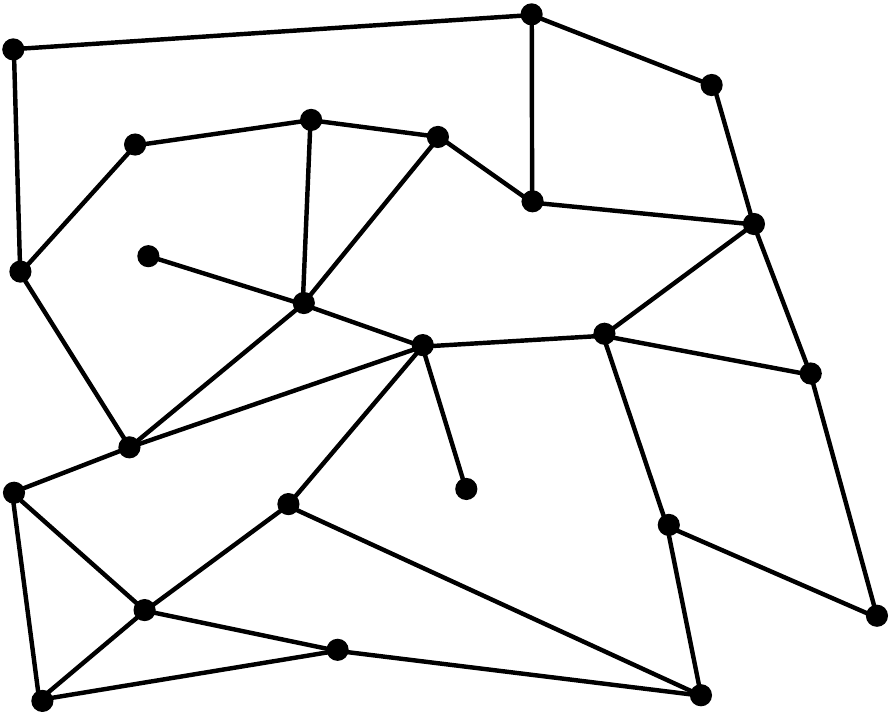}
    \end{center}
    \caption{Realistic urban network.}
    \label{fig: network_big}
\end{figure}

To conclude this section we now confirm the above findings on a more realistic road network. To this end, rather than 
utilising the simple network previously illustrated, we now consider the artificial, but nevertheless realistic, 
network shown in Fig~\ref{fig: network_big}, where we assume that vehicles are allowed to travel in 
both directions on each road. We simulate the traffic flows ustilising the previously mentioned 
simulator SUMO (Simulation of Urban MObility). SUMO has been developed at the Institute of 
Transportation Systems at the German Aerospace Center and is an open source traffic simulation package 
that has been frequently used for large traffic networks. Once pre-defined start and destination roads 
are chosen, SUMO can automatically assign shortest routes to vehicles (e.g., minimum time routes) to 
the vehicles. After the simulation, statistics such as average travel times, average speeds, junction 
turning probabilities are available from SUMO for the whole network, and can be used to form the 
transition matrix $\P$ of travel times. Then, the average speed model can be used to form the 
transition matrix of tire emissions as explained in Section~\ref{tire_Emissions_Model}.\newline 

\begin{figure}[!h]
    \begin{center}
        \includegraphics[width=\textwidth]{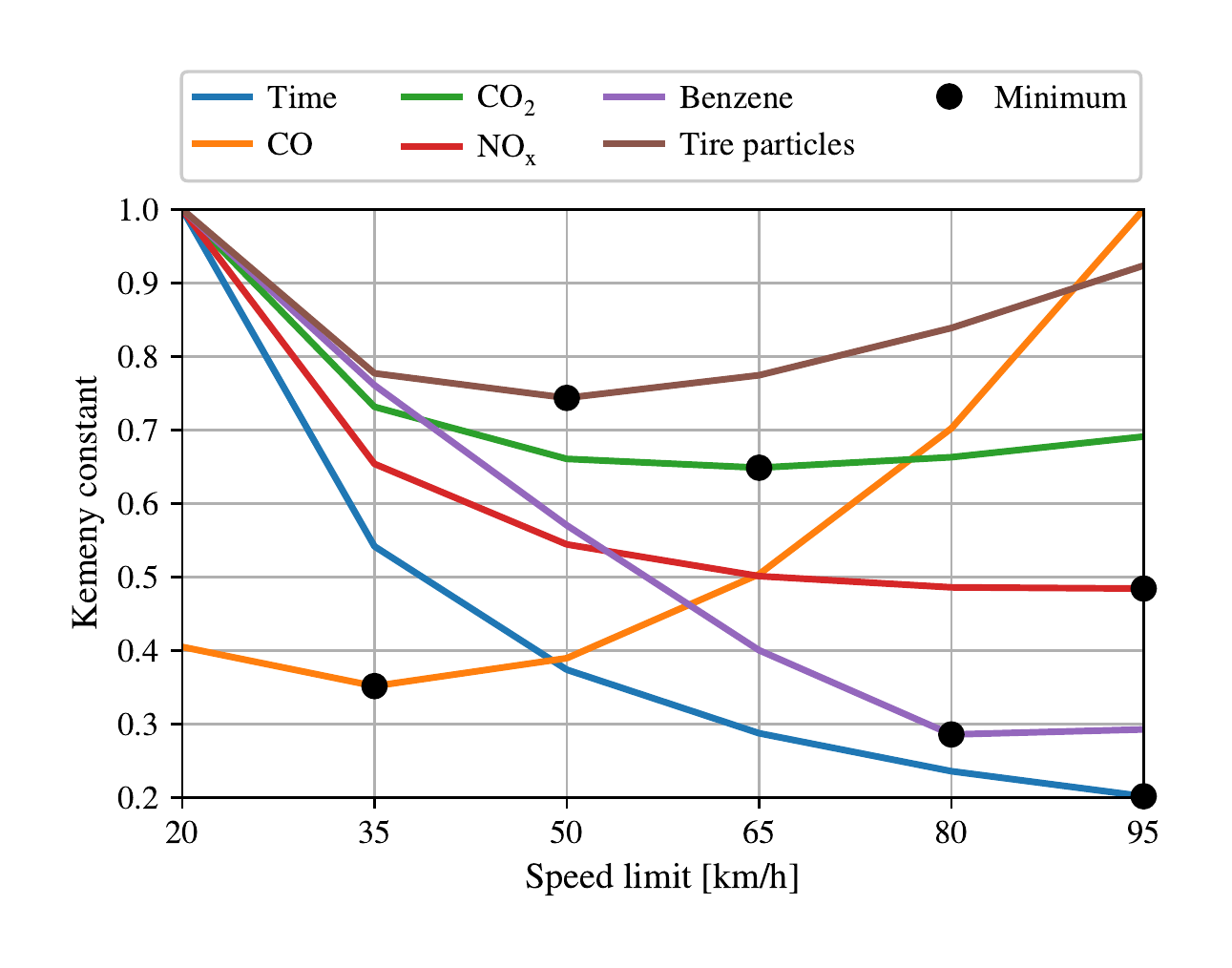}
    \end{center}
    \caption{Kemeny constants for the realistic big network with different speed limits in the 
    entire network.}
    \label{fig: network_big_kemeny}
\end{figure}

As before, in order to calculate the optimal speed limit for the network, simulations for different maximum speeds, which are 
$20, 35, 50,65, 80,95$\,km/h, have been conducted using SUMO. After simulating the network for these six 
different speed limits, and building the resulting Markov chain, six different Kemeny constants can be 
calculated for each type of pollutant as well as for the travel time. Fig~\ref{fig: network_big_kemeny} 
shows the Kemeny constants, normalized to fit the same graph. It can be clearly observed that the optimal 
speed limit varies again for different types of pollutants, confirming the results that had been provided 
for the simpler network. While low speed limits seem to be good for reducing CO and tire particles, high speed limits would be better to reduce $NOx$ and of course travel time. 
\newline

\subsection{Application II - Advisory Systems for Protection of Cyclists}
%\textcolor{red}{We now consider the use of emissions models to develop recommendation systems for road users.}\newline 

\begin{figure}[!h]
    \begin{center}
        \includegraphics[width=\textwidth]{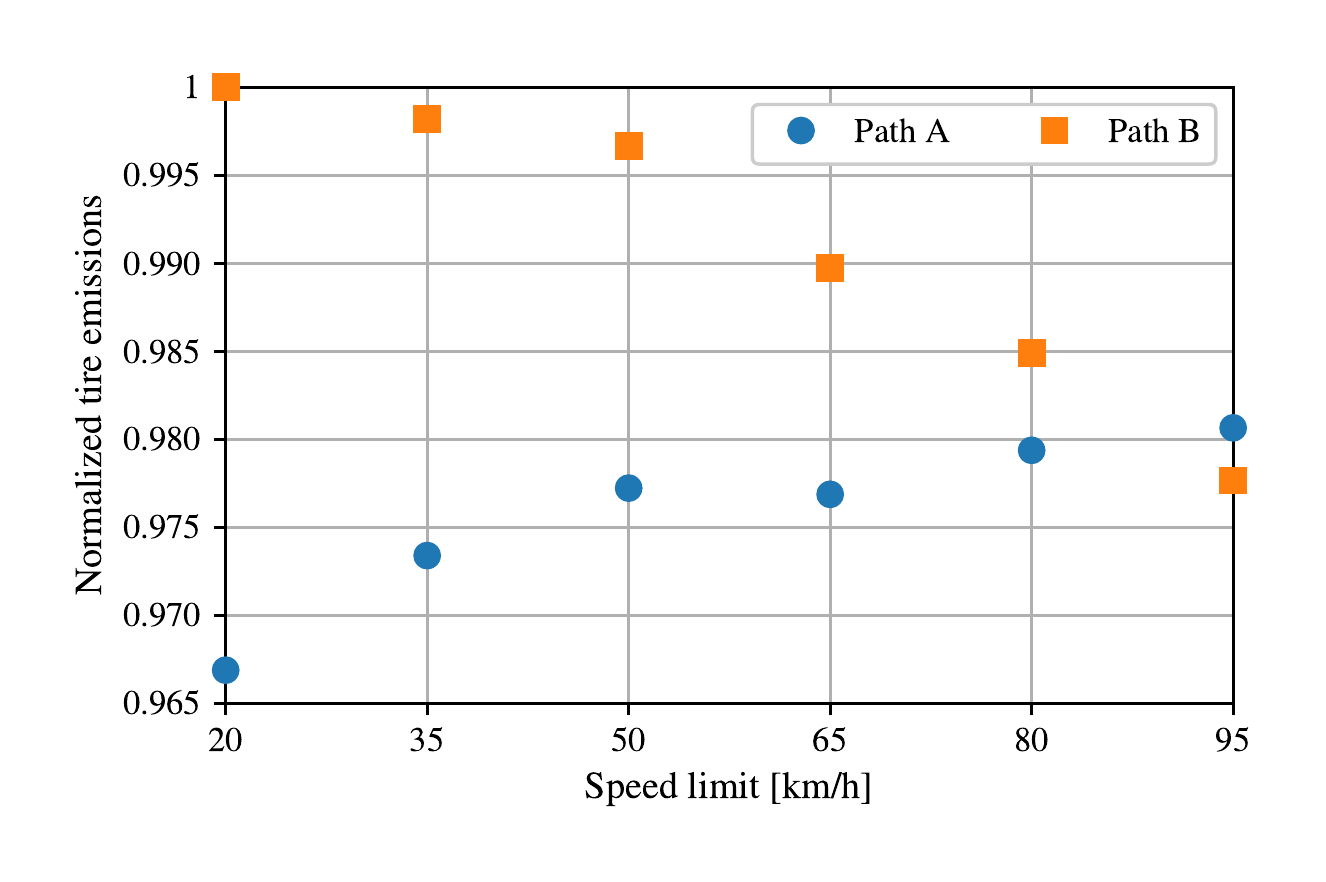}
    \end{center}
    \caption{Minimum tire emission route for the realistic big network.}
    \label{fig: network_big_cyclist}
\end{figure}

One potential application of our Markov chain is related to the tire emissions footprint associated 
with specific routes. Active travel (cycling, walking) is experiencing a resurgence across the developed 
world as citizens abandon public transportation in response to health related concerns associated with 
Covid-19~\cite{Buehler_2021, Combs_2021}. Pedestrians and cyclists are extremely vulnerable road users and their exposure to traffic 
emissions regularly far exceeds that of car occupants.  Given this context our goal now is to use the Markovian 
model to find the minimum tire emission route for cyclists in the network in order to reduce the 
emission exposure and consequently the harmful effect of emissions for their health. Here, we are using 
the classic Dijkstra algorithm~\cite{Dijkstra} to determine the best route, but different from 
traditional applications, we do not wish to minimize distance or time, but the exposure to tires 
emissions. Thus, we associate each road segment with its corresponding entry in the Perron eigenvector 
(which we remind represents the normalized long run fraction of tire emissions release along each road 
segment), and we use Dijkstra algorithm to find the best path. In addition to computing the minimum tire 
emission route, one may also ask whether these best paths are sensitive to changing speed limits; namely, 
in other words, to know whether changing the speed limits in the network also changes the minimum path. 
Fig~\ref{fig: network_big_cyclist} compares the normalized tire emissions along two possible paths, 
having the same specified origin and destination point, for a cyclists, as a function of speed limits. 
In particular, path \textit{A} is the shortest path and the minimum emissions path when speed limits are 
between 20 and 80\,km/h whereas path \textit{B} becomes the minimum tire emission path for higher speed 
limits.\newline 

\subsection{Application III - Tire-Dust Collection}
We now present a somewhat unconventional application of the Markovian approach; namely, using the Markovian approach to inform the collection of tire dust by road sweepers~\cite{Road_Sweepers}. Street sweeping is an effective practice to reduce the amount of road dust, and there is a recent interest in the literature to evaluate the effectiveness of the process~\cite{Polukarova_2020} and to improve the efficiency of sweeping machines~\cite{Korytov_2020}. Here, we take a different view, and we are rather interested in the path followed by road sweepers. Indeed, we have already mentioned that tire particles are harmful to humans irrespective of whether they become airborne, or become part of ground debris. Ground debris is very harmful to humans due to the various pathways for tire particles to enter the human food system; in particular, through city drainage systems. This latter aspect is an important consideration to motivate the collection of tire particles prior to heavy rainfall events, or other severe weather events. In such circumstances, it is important to collect as much tire debris as quickly as possible, and this is in severe contrast to how road sweeping currently takes place.\newline

Typically, road sweepers follow pre-defined paths, trying to cover most of the city, but without taking into account parts that would maximize the collection of tire particles. In this particular context, our Markovian model has much to offer. Our basic intuition is as follows. Since the Perron eigenvector provides the long-run fraction of 
pollutants along each road segment, important information can be extracted from the Markov chain to inform collection of tire 
particles in an optimum manner. However, as we have mentioned - our model is approximate and subject 
to much uncertainty. Thus, we propose to 
use our {\em estimated} chain to seed a learning based algorithm, using Reinforcement Learning (RL), that 
will learn the routes that are most likely to have large quantities of tire particles, and we now indicate how this 
can be achieved making use of the Markovian modelling approach.\newline 

Recall that reinforcement learning~\cite{SuttonRL} is a machine learning 
strategy where agents (such as road sweepers) can explore an unknown %Markovian 
% the RL environments are not necessary Markovian
environment, and learn 
optimal policies (such as the most likely route to find large quantities of tire particles). 
Reinforcement learning in conjuction with our Markovian models is appealing for this problem for 
two reasons.\newline

\begin{itemize}
    \item[(i)] Our Markovian model could, in principle, be used as a basis for routing algorithms. However, 
        as mentioned, the model is very uncertain, as it neglects several factors that affect tire particle generation. 
        Thus, using a learning strategy to tune the elements of the transition matrix to provide a 
        basis for routing makes a great deal of sense for such applications. 
    \item[(ii)] In addition, as we have already mentioned, the Perron eigenvector of the tire 
        emission chain can be used to find the minimum tire emission routes for cyclists and other 
        vulnerable road users. However, the {\em road sweeping} problem is much more challenging if 
        one wishes to find the {\em maximum polluted} routes. This problem is an example of longest path 
        problems and these are known to be {\em NP-hard.} While it is true that longest path problems 
        can sometimes be converted into shortest path problems by negating the edge weights in a 
        graph~\cite{SedgewickAlgorithms}, many shortest path algorithms are able to solve the problem only
        if the underlying graph does not any cycles. This is not common for road network graphs, 
        thus further motivating our interest in reinforcement learning algorithms. 
\end{itemize}

%\textcolor{green}{\bf Well-posed in what sense? } 
To orchestrate a setting for reinforcement learning that is {\em well-posed}, we must first ensure 
that negative cycles\footnote{A negative cycle in a graph is a cycle for which the 
overall sum of the weights is negative.} in the graph associated with the network transition matrix are avoided. To avoid such negative cycles in a graph, we simply add travel time/distance 
constraints to the longest path problem. In addition to making our solution well-posed, such constraints in fact are very sensible for road sweeping 
applications due to limited battery capacity of sweeping vehicles in the case of electric road sweepers. To this end, 
we combine the tire emission graph with a distance graph into one directed weighted graph $G$. 
Recall that an entry of the Perron eigenvector represents the 
normalized long run fraction of tire emissions released along the road segment 
assigned to that entry. The tire emission graph is then obtained by assigning negated entries 
of the Perron eigenvector to the corresponding edges in the graph. The distance graph is derived  
from a road network where the edge weight represents the length of road segments included in the
corresponding state~\cite{SPTokenIEEE}. This 
then turns our problem into a type of multi-objective optimisation problem. We use a convex linear 
combination of the two objectives (travel distance and tire emissions) characterised by a quantity 
$\alpha$ which is the weight of the distance component of the cost. The corresponding weight function 
is described in Function 1 which returns the weight of a given edge in graph $G$. This weight 
represents the cost of traversing state $s$ from any other preceding state.\newline

\begin{algorithm}[h]
    \begin{algorithmic}[1]
        \Require{$\alpha \in \left[0, 1\right]$; $s \in \mathcal{S}$;
            $l_s, c_s, L_{tot}, C_{tot} \in \mathbb{R}$.}
        \Ensure{$w_s$.}
        \Function{$\mathcal{W}$}{$\alpha, s, l_s, c_s$}
            \State
            $w_s = \alpha*l_s +
                \left(1-\alpha\right)*\left(-c_s*\frac{L_{tot}}{C_{tot}}\right)$
            \Statex
            \Return $w_s$
        \EndFunction
    \end{algorithmic}
    \caption*{\textbf{Function 1} The Weight Function\label{alg:weight_func}}
\end{algorithm}

\textit{Notation for the Weight Function}: In Function 1 we have:
$\alpha$ is the distance weight, a real number that satisfies $0 \leq \alpha \leq 1$;
$s$ is a state in the state space $\mathcal{S}$;
$l_s$ is the length of road links included in state $s$\footnote{The road merging 
mechanism introduced in~\cite{SPTokenIEEE} is utilised.};
$c_s$ is the amount of tire particles emitted along the road segments included in state $s$ 
within some given time interval $\tau$;
$L_{tot}$ is the sum of the road lengths for a given road network;
$C_{tot}$ is the total amount of tire particles emitted through the entire road network 
within time interval $\tau$. $C_{tot}$ can be estimated using some historical data,
and we simply assume that $C_{tot}$ is known.
\newline

To further elaborate on our proposed algorithm, we denote by $\alpha_{min}$ the minimum value of $\alpha$ such that the graph $G$ does not 
have any negative cycles for weights computed as $\mathcal{W}(\alpha, s, l_s , \bar{c}_s)$,
where $\bar{c}_s$ is the estimated amount of tire emissions along the road segments clustered 
in state $s$ within the time interval $\tau$. The value of $\bar{c}_s$ is the result 
of multiplication of $C_{tot}$ by 
the corresponding to state $s$ entry of Perron eigenvector. The value of $\alpha_{min}$ is determined
empirically to two decimal digits of precision. Note that even though the weight $w_s$ defined 
in Function 1 is measured in the units of $l_s$ (i.e., distance), the values of 
$w_s$ can be negative. This particular design of $w_s$ results in a lower value of 
$\alpha_{min}$ compared to what one would obtain in the case of normalized unitless weights.
\newline

Once a {\em combined} graph (without any negative cycles) has been constructed, we then use a shortest path 
algorithm to compute {\em default solutions}. To deal with values of $\alpha$ such that 
$\alpha < \alpha_{min}$, approximation techniques are required to find the maximum tire 
emissions routes subject to the distance
constraints. To solve this problem, our preferred approximation tool is reinforcement learning. The 
{\em default solutions} from the Markov chain are used as the initial estimate for the reinforcement learning algorithm. 
Note that even though the number of tire particles is generally larger along longer routes, 
assigning very long routes to the road sweepers would dramatically increase their travel time and may 
not even be feasible due to battery constraints.
\newline
 
We employ reinforcement learning to amend the initial 
estimate whenever the constraint of $\alpha < \alpha_{min}$ is satisfied (note that the case 
$\alpha = 1$ corresponds to the shortest path routing).
The Q-learning algorithm proposed in~\cite{EvanDar}
is utilised here in our work. The initial parameters~\cite{SuttonRL} for the underlying Q-learning 
algorithm are obtained from the default solutions of the Markovian model. Actions of the agent 
(i.e. road sweeper) represent 
road directions, for instance, turn left/right. The goal of the agent is to find a route which maximises
the total expected reward~\cite{SuttonRL}.
The reward function for this application is outlined in 
Function 2: it returns a reward at time step $t$.\newline

\begin{algorithm}[h]
        \begin{algorithmic}[1]
            \Require{$\alpha \in \left[0, 1\right]; t, H \in \mathbb{N}; s, s_{D} \in \mathcal{S};
                \beta_1, \beta_2 \in \mathbb{R}^+$.}
            \Ensure{$r_t$.}
            \Function{$\mathcal{R}$}{$s$}
                \If {$s \neq s_D$} \Comment{{\color{blue}Destination not reached yet}}
                    \If {$t \neq H$} \Comment{{\color{blue}Time horizon not reached yet}}
                        \State
                        Get the length $l_s$ of road links included in state $s$.
                        \State
                        Get the estimated amount of pollution $\bar{c}_s$ in state $s$.
                        \State
                        Measure the actual amount of pollution $c_s$ emitted in state $s$.
                        \Statex
                        \textit{{\color{blue}\ \ \ \ \ \ \ \ \ \ \ \ \ 
                            // Compute weights for state $s$}}
                        \State
                        $\bar{w}_{s} = \mathcal{W}(\alpha, s, l_s, \bar{c}_s),$\,\,
                        $w_{s} = \mathcal{W}(\alpha, s, l_s, c_s)$
                            \Comment{{\color{blue}Calls to Function 1}}
                        \State
                        $r_t = 1 - \frac{\bar{w}_s}{w_s}$
                    \Else \Comment{{\color{blue}Time horizon reached}}
                        \State
                        $r_t = -\beta_2$
                            \Comment{{\color{blue}Penalty}}
                    \EndIf
                \Else \Comment{{\color{blue}Destination reached}}
                    \State
                    $r_t = \beta_1$
                \EndIf
                \Return $r_t$
            \EndFunction
    \end{algorithmic}
    \caption*{\textbf{Function 2} The Reward Function\label{alg:reward_func}}
\end{algorithm}

\textit{Notation for the Reward Function:} The notation for Function 2 is as follows. 
$H$ is the time horizon, i.e.\ the number of allowed transitions per episode (day). 
State $s_D$ is the destination state. 
The parameter $\beta_1$ represents a reward that an agent would receive if it reached the 
destination state.
Finally, $\beta_2$ is a reward given to the agent if it did not reach the destination 
within $H$ times steps. 
\newline

A realistic road network, based on an existing area in Barcelona, Spain, used in all our experiments 
is depicted in Fig~\ref{fig:network_barcelona}.
To illustrate our algorithm we describe several illustrative experiments, designed using the SUMO traffic simulator and 
randomly generated traffic conditions.
In all our experiments, a single Q-learning agent, i.e.\ road sweeper, starting from the same origin O 
(see Fig~\ref{fig:network_barcelona}) each episode (day) is used. The agent has a fixed destination 
D to which it is asked to find the optimal route which (i) should be most polluted, and (ii) satisfies 
the distance constraint given by the parameter $\alpha$. A road sweeper is released every time a new 
episode starts, i.e., once per day. Regarding the design parameters in the reward function, the values 
of $\beta_1$ and $\beta_2$ were tuned to be $3$ and $8$, respectively. The values of $\alpha_{min}$
were empirically determined for each specific experiment.\newline

\subsubsection{Experiment 1: Tire dust collection under high traffic densities}
In this experiment, we firstly consider a scenario in which traffic density is high. To achieve such 
a condition, we release a new vehicle every simulation step. High traffic density conditions naturally 
result in a larger amount of tire dust on the streets. In these settings, the estimated value of 
$\alpha_{min}$ is $0.78$.
\newline

\begin{figure}[!h]
    \centering
    \includegraphics[width=0.8\textwidth]{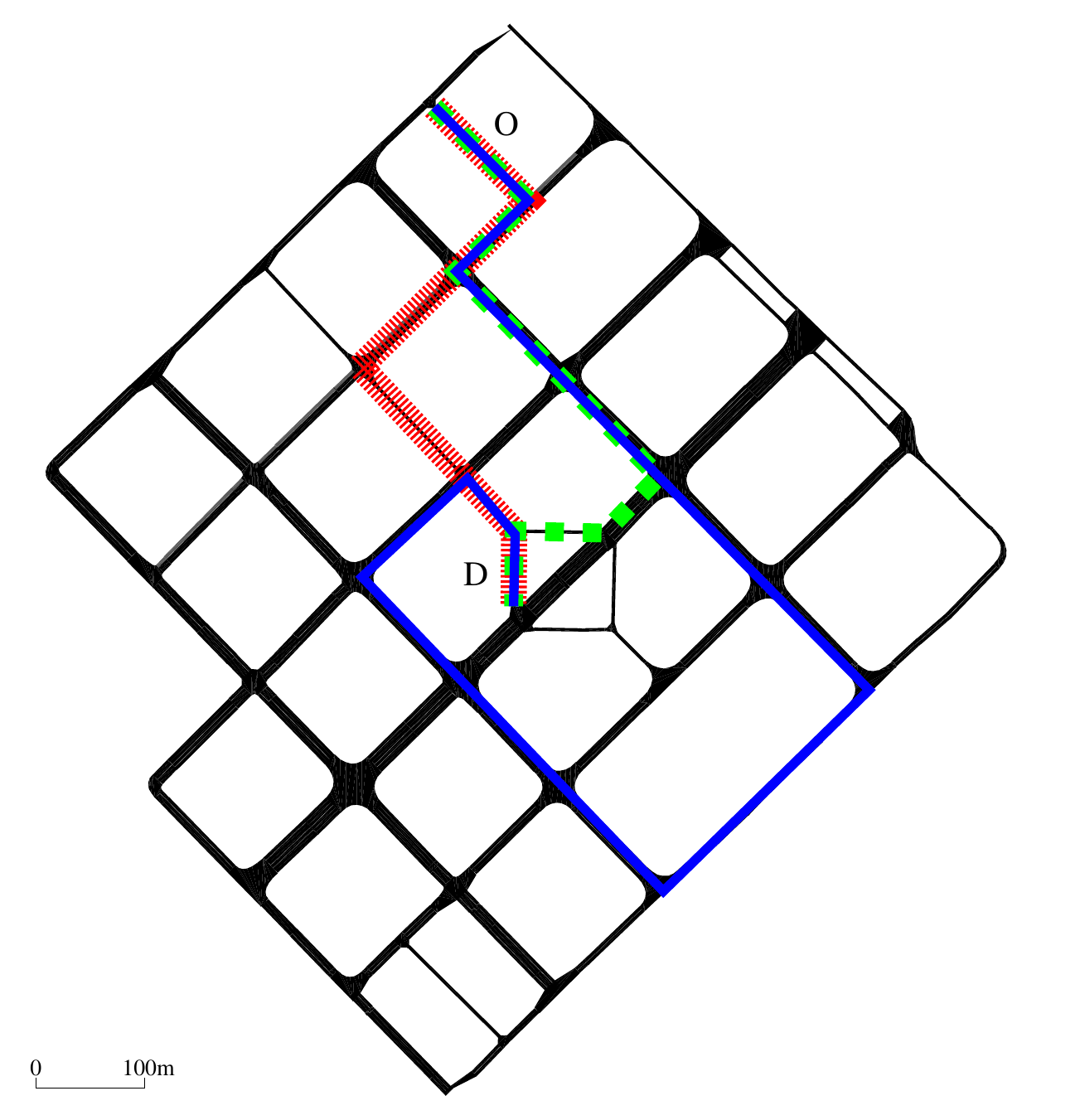}
    \caption{Realistic road network used in the experiments: an area in Barcelona, Spain. 
        The road networks includes 153 road links which were clustered in 62 states. Fixed origin
        O and destination D are used in the experiments. Note: the route marked 
        with red is the shortest path route from O to D; the green route is the default solution; and 
        the blue one represents a solution provided by the Q-learning algorithm under 
        high traffic density.
    }\label{fig:network_barcelona}
\end{figure}
\begin{figure}[!h]
    \centering
    \includegraphics[width=0.8\textwidth]{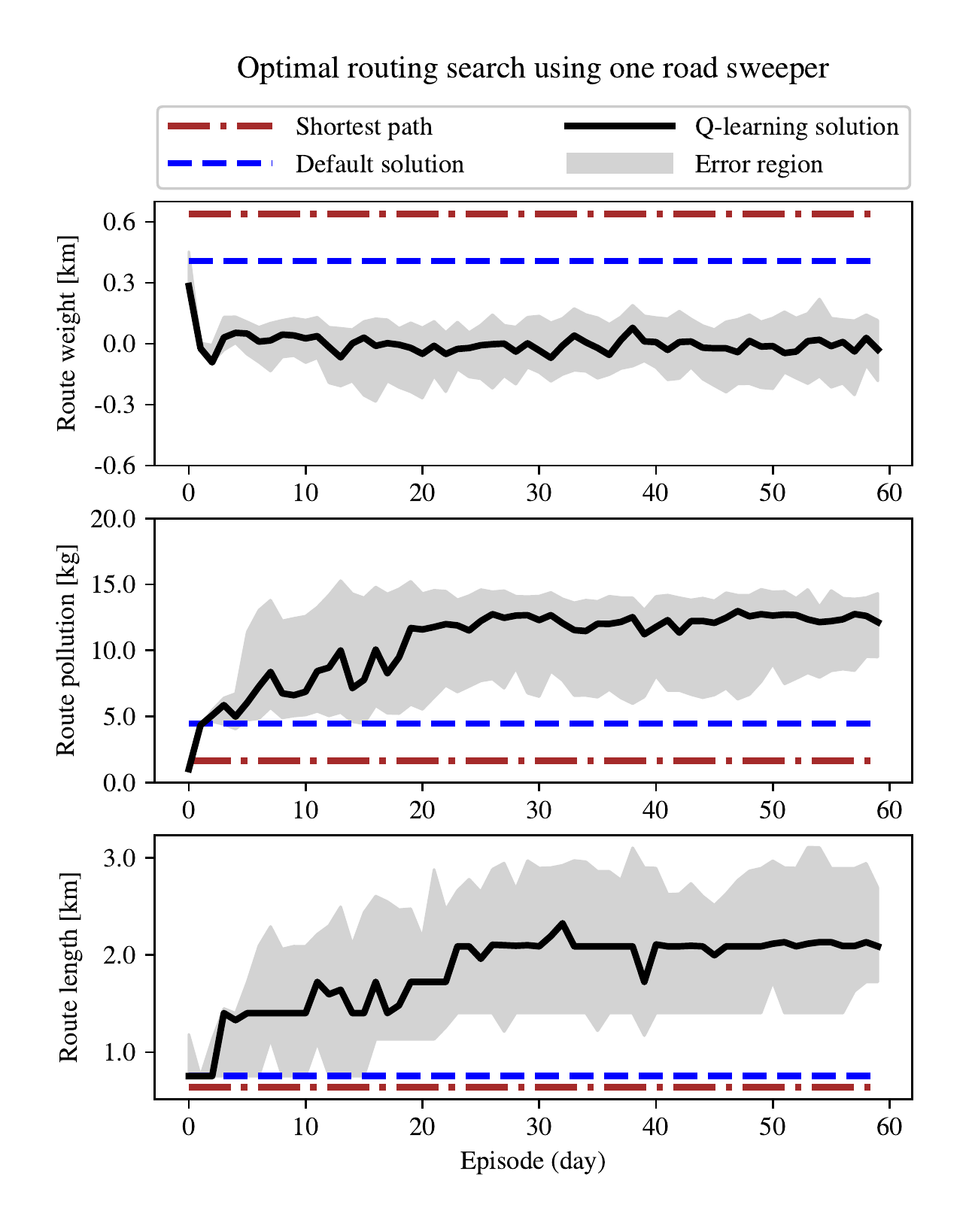}
    \caption{Comparison of the shortest path routing, the initial solution ($\alpha_{min}=0.78$) and 
        Q-learning solution for $\alpha=0.5$ under high density traffic conditions. 
        The black curve corresponds to the median value 
        of 100 different realizations of the experiment. The error region indicates 
        the 30th and 70th percentiles.}\label{fig:network_barcelona_qlearning_high}
\end{figure}

Fig~\ref{fig:network_barcelona} 
shows the shortest path in red, the initial solution in dashed green, and the optimal solution in 
solid blue, obtained using the proposed Q-learning algorithm for high traffic densities. 
Fig~\ref{fig:network_barcelona_qlearning_high} compares the properties of such three routes with 
$\alpha=0.5$ used for the Q-learning solution. 
The brown dot-dashed line corresponds to the shortest path route, which was calculated using Dijkstra 
shortest path algorithm on the graph with weights computed using Function 1 for $\alpha=1$. The blue 
dashed line corresponds to the route obtained from the initial estimate, i.e., the default solution. 
Such a route was also calculated using Dijkstra shortest path algorithm on a graph with weights computed 
using the same weight function for $\alpha=\alpha_{min}=0.78$. As it can be seen in 
Fig~\ref{fig:network_barcelona_qlearning_high} (black solid line), the road sweeper uses the initial 
solution at the beginning of the learning process. The agent, however, can explore the environment by 
taking random actions and eventually improves the chosen sweeping path. As it can be observed, the agent 
is able to find much longer routes than the shortest path and default routes in a rapid fashion, and 
such routes are also more polluted with tire particles. Even though routes with higher tire emissions 
have been explored by the agent, it will not prefer them due to the distance constraint, and this 
Q-learning routing system eventually converges to the solution which is optimal for the selected value 
of $\alpha$.
\newline

\subsubsection{Experiment 2: Tire dust collection under low traffic densities}
\begin{figure}[!h]
    \centering
    \includegraphics[width=0.8\textwidth]{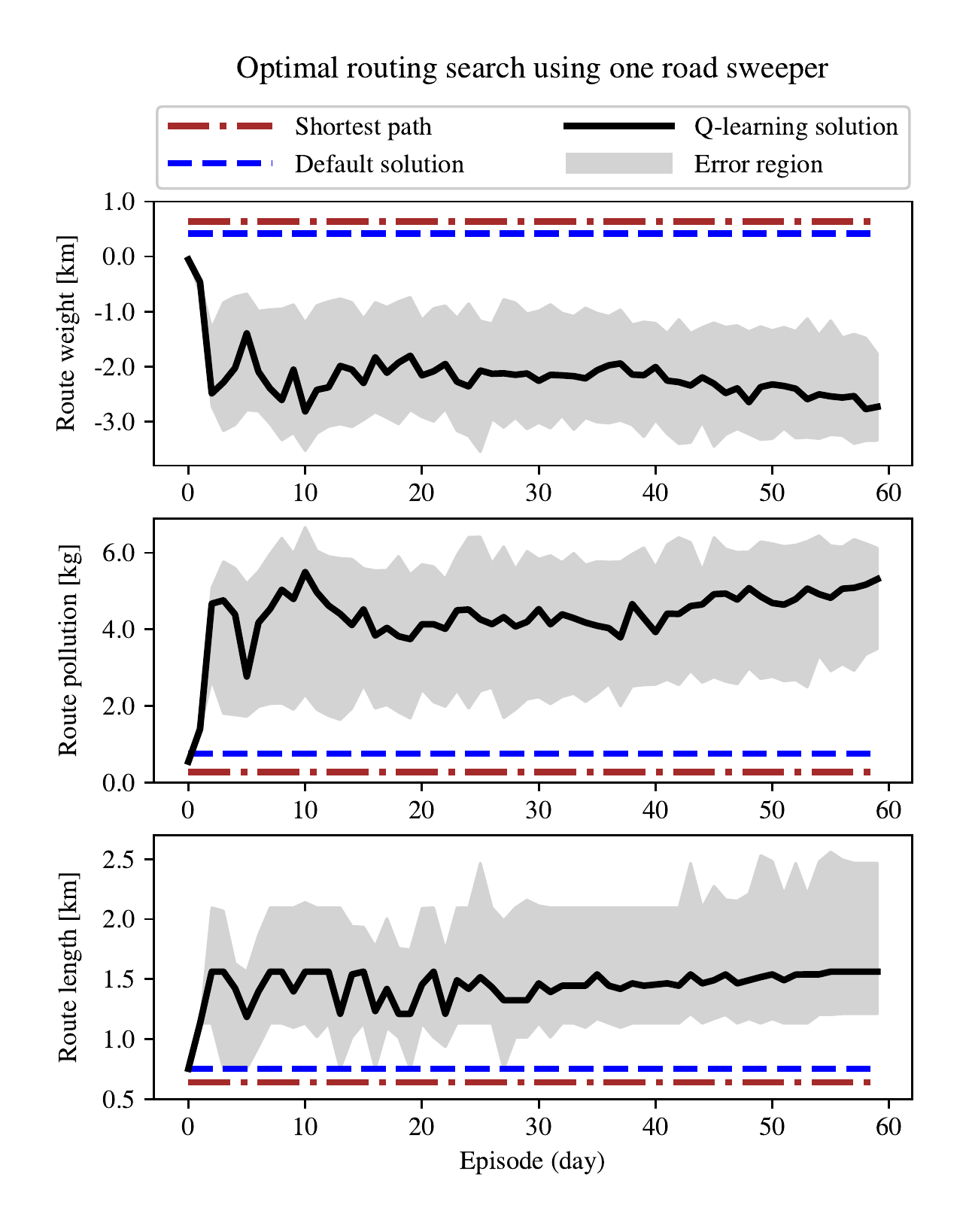}
    \caption{Comparison of the shortest path routing, the default solution ($\alpha_{min}=0.79$) and 
        Q-learning solution for $\alpha=0.4$ under low density traffic conditions. 
        The black solid curve is the median value 
        of 100 different realizations of the experiment. The error region indicates 
        the 30th and 70th percentiles.}\label{fig:network_barcelona_qlearning_low}
\end{figure}

To simulate traffic conditions with a lower density, a new vehicle is now released every second 
simulation step. In this case, the obtained value of $\alpha_{min}$ is $0.79$.
Fig~\ref{fig:network_barcelona_qlearning_low} depicts the shortest path, default solution, and 
performance of Q-learning for $\alpha = 0.4$. For low densities, it is reasonable to reduce 
the value of $\alpha$ in order to give more priority to pollution over distance. 
Otherwise, the Q-learning routing system 
would be useless as it may converge to the default solution. Note that the amount of collected pollution 
by the road sweeper in this case is indeed lower than that in the case of high density traffic 
(see the middle subplots in Fig~\ref{fig:network_barcelona_qlearning_high} and 
Fig~\ref{fig:network_barcelona_qlearning_low}).\newline

From Experiment 1 and Experiment 2, we can draw the conclusion that the system is indeed able to find 
the optimal solution (the most polluted route with tire particles, without breaking travel 
distance/time constraints) in both high and low density traffic conditions.
Thus, the RL strategy is an attractive alternative 
to those tools that can be fragile, especially when dealing with longest path problems and 
large-scale scenarios.

\section*{Conclusion}
The problem of tire dust collection is likely to become one of the most pressing issues 
in automotive research and in wider society. While the problem of micro-plastic pollution is already becoming 
as issue of concern, the problem of tire induced pollution has, remarkably, yet to manifest 
itself in the consciousness of the public-at-large, possibly due to the sheer weight of the 
zero-tailpipe narrative that prevails currently in public discourse. Our objective in this 
paper is thus twofold. First, we wish to make researchers across a wide spectrum of 
disciplines, aware of this problem, in all its guises. Second, we wish to suggest 
mitigation measures that can be used to combat this problem. While previous 
studies have focussed on {\em on-vehicle} mitigation measures, and 
network level {\em access control mechanisms}, our approach here is somewhat different. 
Our approach is to develop modelling strategies that can be deployed {\em a-posteri}. Specifically, 
we wish to predict, using a combination of measurements, and analytics, the likely areas 
where tire-dust will aggregate, with a view to using this information to inform collection strategies. 
In this paper we have introduced one such model of tire dust distribution in cities. A number of 
application use-cases are suggested that use the main features of this model. Future work 
will explore refinements of this initial model and its experimental validation.

%\section*{Acknowledgments}
%This paper is supported by TUBITAK (The Scientific and Technological Research Council of Turkey).

\nolinenumbers

% Either type in your references using
% \begin{thebibliography}{}
% Text
% \end{thebibliography}
%
% or
%
% Compile your BiBTeX database using our plos2015.bst
% style file and paste the contents of your .bbl file
% here. See http://journals.plos.org/plosone/s/latex for 

\end{document}